\documentclass{article}
\usepackage{spconf,amsmath,graphicx}
\usepackage{caption,array,booktabs,multirow,cite}
\usepackage[dvipsnames]{xcolor}

\newcolumntype{C}[1]{>{\centering\let\newline\\\arraybackslash\hspace{0pt}}m{#1}}
\title{Breast Lesion Segmentation in Ultrasound Images with Limited Annotated Data}
\name{Bahareh Behboodi$^{\star \dagger}$ \qquad Mina Amiri$^{\star \dagger}$ \qquad Rupert Brooks$^{\star \ddagger}$ \qquad Hassan Rivaz$^{\star \dagger}$}

\address{$^{\star}$Department of Electrical and Computer Engineering, Concordia University, Canada  \\ $^{\dagger}$PERFORM Center, Concordia University, Canada \\ $^{\ddagger}$Nuance Communications}

\begin{document}
%
\maketitle
\begin{abstract}
Ultrasound (US) is one of the most commonly used imaging modalities in both diagnosis and surgical interventions due to its low-cost, safety, and non-invasive characteristic. US image segmentation is currently a unique challenge because of the presence of speckle noise. As manual segmentation requires considerable efforts and time, the development of automatic segmentation algorithms has attracted researchers' attention. Although recent methodologies based on convolutional neural networks have shown promising performances, their success relies on the availability of a large number of training data, which is prohibitively difficult for many applications. Therefore, in this study we propose the use of simulated US images and natural images as auxiliary datasets in order to pre-train our segmentation network, and then to fine-tune with limited \textit{in vivo} data. We show that with as little as $19$ \textit{in vivo} images, fine-tuning the pre-trained network improves the dice score by $21\%$ compared to training from scratch. We also demonstrate that if same number of natural and simulation US images is available, pre-training on simulation data is preferable.
\end{abstract}
\section{Introduction}
\label{sec:intro}
Breast cancer has been reported as one of the leading causes of death among women worldwide. Although, digital mammography is an effective modality in breast cancer detection, it has limitations in detecting dense lesions which are similar to dense tissues \cite{ref2}, and further uses ionizing radiation. Therefore, ultrasound (US) imaging as a safe and versatile screening and diagnostic modality plays an important role in this regard. However, due to contamination of the US images with speckle noise, US images have low resolution and poor contrast between the target tissue and background; thus, their segmentation is currently a challenging task \cite{ref10}. Researchers have utilized recent state-of-the-art deep learning techniques in order to overcome limitations in manual segmentation. Despite the success of deep learning techniques in computer vision tasks, their performance depends on the size of input data which is limited specially in medical US images. The collection and annotation of US images require considerable effort and time which attain the need to a deep learning-based strategy that can be trained on as few annotated data as possible. 

The U-Net architecture \cite{ref1}, as one of the most well-known networks for segmentation purposes, is built upon fully convolutional network. It involves several convolutional, max-pooling, and up-sampling layers. To cope with limited input data for training U-Net, researches have proposed various strategies based on data augmentation and transfer learning\cite{ref10,ref11,ref12}. Data augmentation cannot truly capture the characteristics of the real data when very limited data is available. To this end, we propose a methodology based on transfer learning which utilizes US simulation data and natural images as an auxiliary dataset. The goal is to enhance the segmentation results while only few images are available. In our work, first, we pre-train the U-Net with US simulated and natural images separately, and then fine-tune the network with only $15\%$ of available \textit{in vivo} images. We demonstrate $21\%$ improvement in segmentation results when small number of images are available.

\section{Method}
\label{sec:intro}
\begin{figure}[h!]
  \centering
  \centerline{\includegraphics[width=6cm]{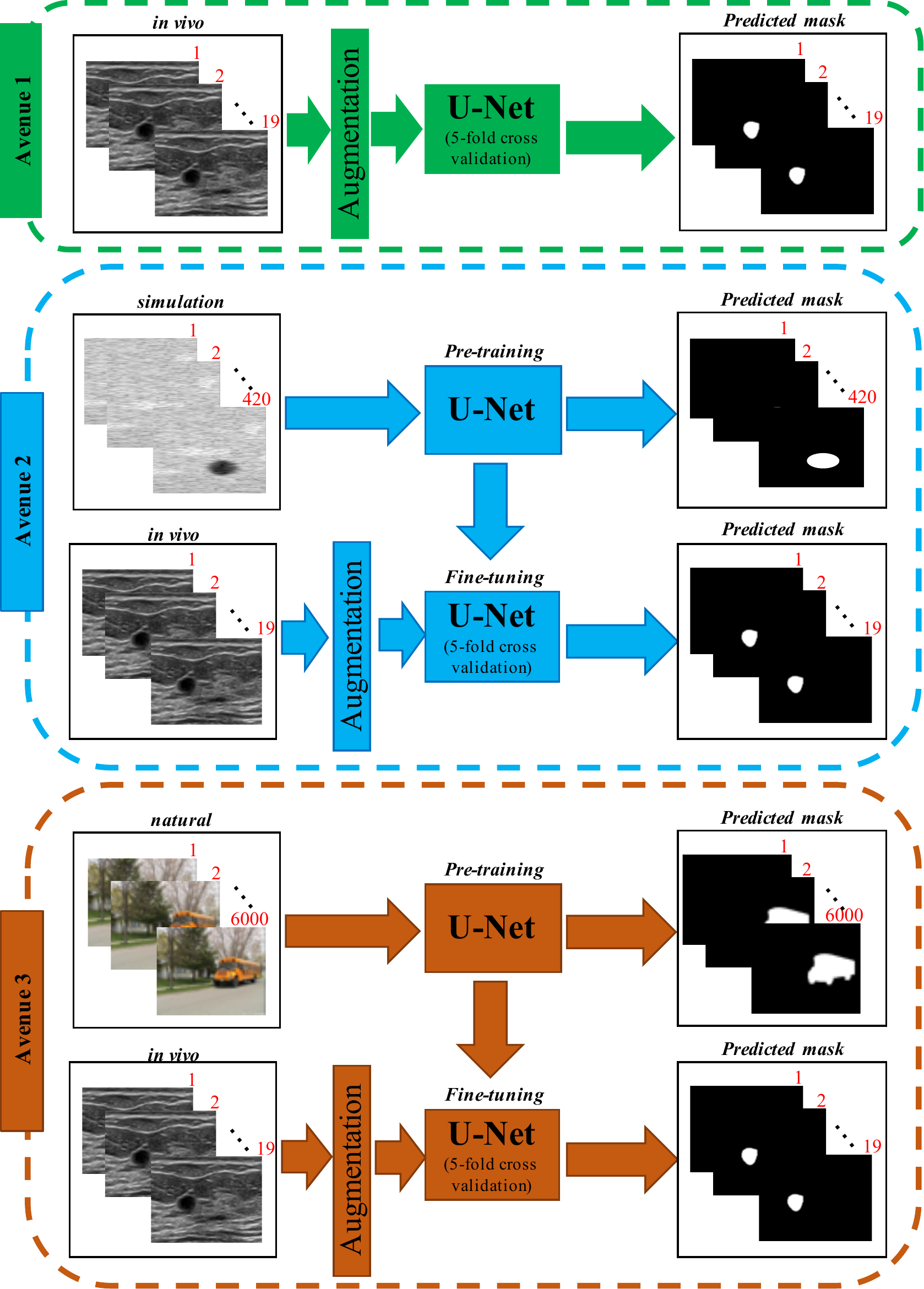}}
  \caption{Proposed workflow for training U-Net when limited annotated data is available.}
  \label{fig:workflow}
\end{figure}

In deep learning approaches, the improvement in results depends on the number of training data. Therefore, such techniques perform better if they have larger amount of training data. In medical images, especially in US images, annotating enough number of training data is expensive, and thus, we take advantage of simulated US data as well as natural images as the auxiliary datasets for pre-training U-Net in our proposed workflow. To that end, our proposed workflow consists of three avenues as shown in Fig. \ref{fig:workflow}. In the first avenue, U-Net is trained using only $15\%$ of the \textit{in vivo} dataset. In the second avenue, U-Net is first pre-trained on the simulated data, and then fine-tuned using the same $15\%$ of the {\it in vivo} dataset which was used in the first avenue. And the last avenue is similar to the second avenue with the difference that natural images were used for pre-training. Section \ref{ssec:exp} will clarify each avenue in more details.

\subsection{\textbf{\textit{In vivo}} Data}
\label{ssec:invdata}
{\it In vivo} dataset includes 163 breast B-mode US images with lesions and the mean image size of $760\times570$. The images as well as their delineation of lesions are publicly available upon request \cite{ref2}. The breast lesions of interest are generally hypoechoic (i.e. tissues with lower echogenicity), that is, darker than surrounding tissue. Only $15\%$ of the total number of {\it in vivo} images were used as training and validation datasets and the remaining $85\%$ were set as the testing datasets. The size of training dataset was selected 4 times larger than the size of validation images yielding $19$, $5$, and $139$ images for training, validation, and testing datasets, respectively. 

\subsection{Simulation Data}
\label{ssec:simdata}
To simulate B-mode images, a MATLAB-based publicly available US simulation software, Field\_II \cite{ref3} was used. Number of RF lines, centre frequency, sampling frequency, and speed of the sound were respectively set to $50$, $3.5 MHz$, $100 MHz$, and $1540 m/s$. In our simulation phantom, the surface started at $30 mm$ from the transducer surface and the axial, lateral, and elevational distances were initiated as $60mm$, $40 mm$, and $10 mm$, respectively. The scatterers were randomly distributed in our virtual phantom such that each $mm^3$ of phantom had in average $4$ scatterers, to allow for fast ultrasound simulation. In our simulated images we considered each image to randomly have either hyperechoic (i.e. tissues with higher echogenicity), that is brighter than surrounding tissue, or hypoechoic lesions, or both at the same time in order to let our network learn better the various possible textures of the US images. The intensities for hyperechoic lesions were set $k$ times higher than the background where $k$ was an integer in range of $1-10$, however, for hypoechoic lesions the intensities were set $l$ times the background where $l$ was a random variable between $0$ and $1$. The location of the lesions was randomly selected with circle or ellipsoid shapes. A total of 700 images were simulated and separated to training, validation, and testing sets with splitting factors of $60\%$, $15\%$, and $25\%$ of total number of images, yielding $420$, $105$, and $175$ images, respectively. It worth mentioning that as the {\it in vivo} data consisted of hypoechoic lesions, in the masks of simulated data only the pixels inside of the hypoechoic lesions were set to $1$, and the remaining pixels were set to $0$. Therefore, there were some simulated images with zero segmented lesions in their masks.

\subsection{Natural Data}
\label{ssec:natdata}
The natural images are publicly available at \cite{ref8}. The dataset consists of $10000$ images of salient objects with their annotations. In our work, the dataset was split to training, validation, and testing sets with splitting factors of $60\%$, $15\%$, and $25\%$ of total number of images, yielding $6000$, $2500$, and $1500$ images, respectively.

\subsection{U-Net Architecture}
\label{ssec:unet}
\begin{figure}[h!]
  \centering
  \centerline{\includegraphics[width=6cm]{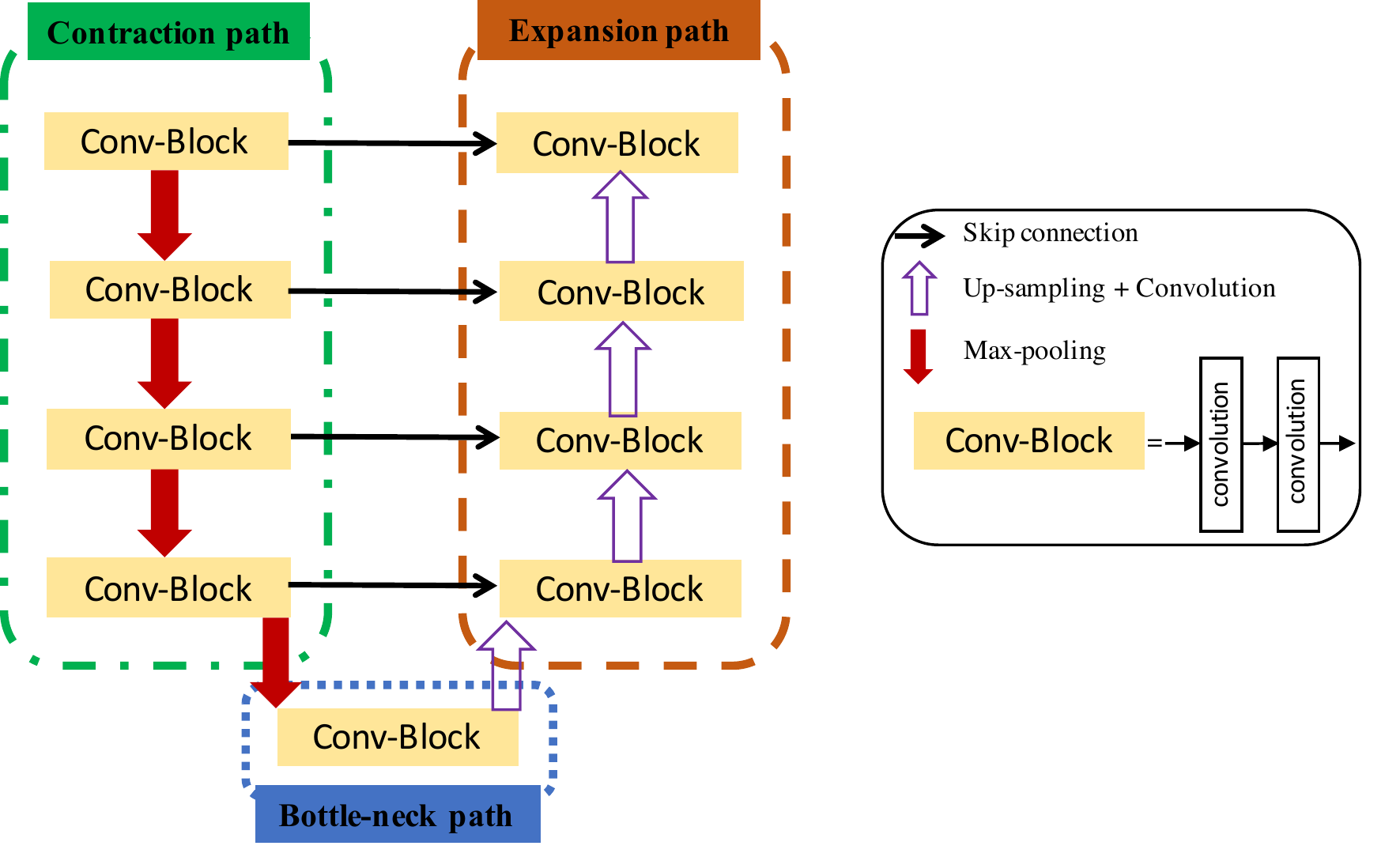}}
  \caption{U-Net structure with its contraction, bottle-neck, and expansion paths.}
  \label{fig:unet}
\end{figure}

The U-Net structure previously proposed by \cite{ref1} utilizes several conv-block, max-pooling, up-sampling, and skip connection layers as illustrated in Fig. \ref{fig:unet}. Each conv-block consists of repetition of two convolution layers while in the contraction and expansion paths, followed by max-pooling and up-sampling layers, respectively. In this work, the kernel sizes of convolution, max-pooling, and up-sampling layers were set to 3$\times$3, 2$\times$2, and 2$\times$2, respectively. As a pre-processing step, all the images were resized to $388\times388$, mirrored with the mirroring factor of $92$ pixels, yielding images with size $572\times572$, and normalized to the range of $[0,1]$. Thus, the size of the input and output data was $(batch, 572, 572, 1)$ and $(batch, 388, 388, 2)$, respectively, where $batch$ indicates the number of images in each batch. The activation and loss functions, optimizer, learning rate, number of epochs, batch size, weight initializer, and kernel regularizer were initialized as stated in Table \ref{tab:unetparam}. The Dice score is defined as $DSC = \frac{2 |G\cap P|}{|G|+|P|}$, where $G$ and $P$ is ground truth and predicted masks, respectively.
\begin{table}[h]
	\centering
	\caption{U-Net parameters}
	\label{tab:unetparam}
	\small
	\begin{tabular}{ m{5cm}  C{2cm}}
		\toprule[0.5mm]
		\multicolumn{1}{c}{\bf Parameter} & \multicolumn{1}{c}{\bf Value} \\
		\toprule
		Activation function (except last layer)		& ReLU \cite{ref5}\\
		Activation function (last layer)		& Softmax \\
		Loss function		& Dice score\\
		Optimizer		& Adam\cite{ref6}\\
		Learning rate	& $0.00001$\\
		No. of epochs	& $150$\\
		Batch size 	& 8\\ 
		Weight initializer	& He-normal\cite{ref7}\\ 
		Kernel-regularizer	& L2-norm\\ 
		\toprule
	\end{tabular}
\end{table}
\subsection{Experiments}
\label{ssec:exp}
As previously mentioned, in this work we propose three avenues to study the impact of simulated and natural imagesas the auxiliary datasets for US segmentation (see Fig. \ref{fig:workflow}). In the following paragraphs, we will explain each avenue in detail:
\subsubsection{Avenue 1: Train U-Net on {\it in vivo} data from scratch}
\label{sssec:av1}
In the first avenue, the U-Net structure with above-mentioned parameters, was trained on {\it in vivo} images from scratch using $19$ and $5$ images as training and validation sets, respectively, and was tested on $139$ images. We call this trained network as {\it Pt\_invivo}. Due to small number of training data, we used $5$-fold cross-validation to prevent variation in performance. Prior to each optimization iteration, we performed "on-the-fly" augmentation by applying random height-shift, width-shift, and zooming.
\subsubsection{Avenue 2: Train U-Net with US simulated images and fine-tune with {\it in vivo} data}
\label{sssec:av2}
In this avenue, U-Net was first trained using $420$ and $105$ simulation images as its training and validation sets, respectively. Similar to the first avenue, the U-Net was initialized using parameters mentioned in Table \ref{tab:unetparam}. For the simplicity, we refer the trained U-Net with simulated data as {\it Pt\_sim}. Afterwards, the contraction path of {\it Pt\_sim} was fine-tuned on {\it in vivo} training and validation sets based on parameters in Table \ref{tab:unetparam} except that weights were initialized using the {\it Pt\_sim} weights. We call the fined-tuned network as {\it Ft\_sim\_invivo} which was tested on {\it in vivo} test set. $5$-fold cross-validation and "on-the-fly" augmentation was used for fine-tuning our {\it Ft\_sim\_invivo} network.
\subsubsection{Avenue 3: Train U-Net with natural images and fine-tune with {\it in vivo} data}
\label{sssec:av3}
In this step, similar to the $Avenue$ $2$ described above, U-Net was first pre-trained and then fine-tuned on {\it in vivo}. However, for pre-training the network we used $6000$ and $2500$ natural images as training and validation sets, respectively. For simplicity, the pre-trained U-Net with natural images is referred as {\it Pt\_nat} and the fine-tuned network using {\it Pt\_nat} is referred as {\it Ft\_nat\_invivo}. $5$-fold cross-validation and "on-the-fly" augmentation was used in the fine-tuning step.

\section{Results}
\label{sec:res}
\subsection{Evaluation Criteria}
In this work, we used Dice Similarity Coefficient ($DSC$) as our evaluation criteria. It is worth noticing that we also used $DSC$ as our loss function, however, in the evaluation step the predicted masks which were the output of the last layer (i.e. Softmax layer) were first binarized using $argmax$ function, and then compared with the ground truth masks. As our dataset was unbalanced (i.e. number of background pixels were higher than the lesion pixels), we only report the $DSC$ scores of the foreground (i.e. lesions) masks ignoring the $DSC$ score of the background. 
\subsection{Experimental Results}
\begin{figure}[h!]
  \centering
  \centerline{\includegraphics[width=8cm]{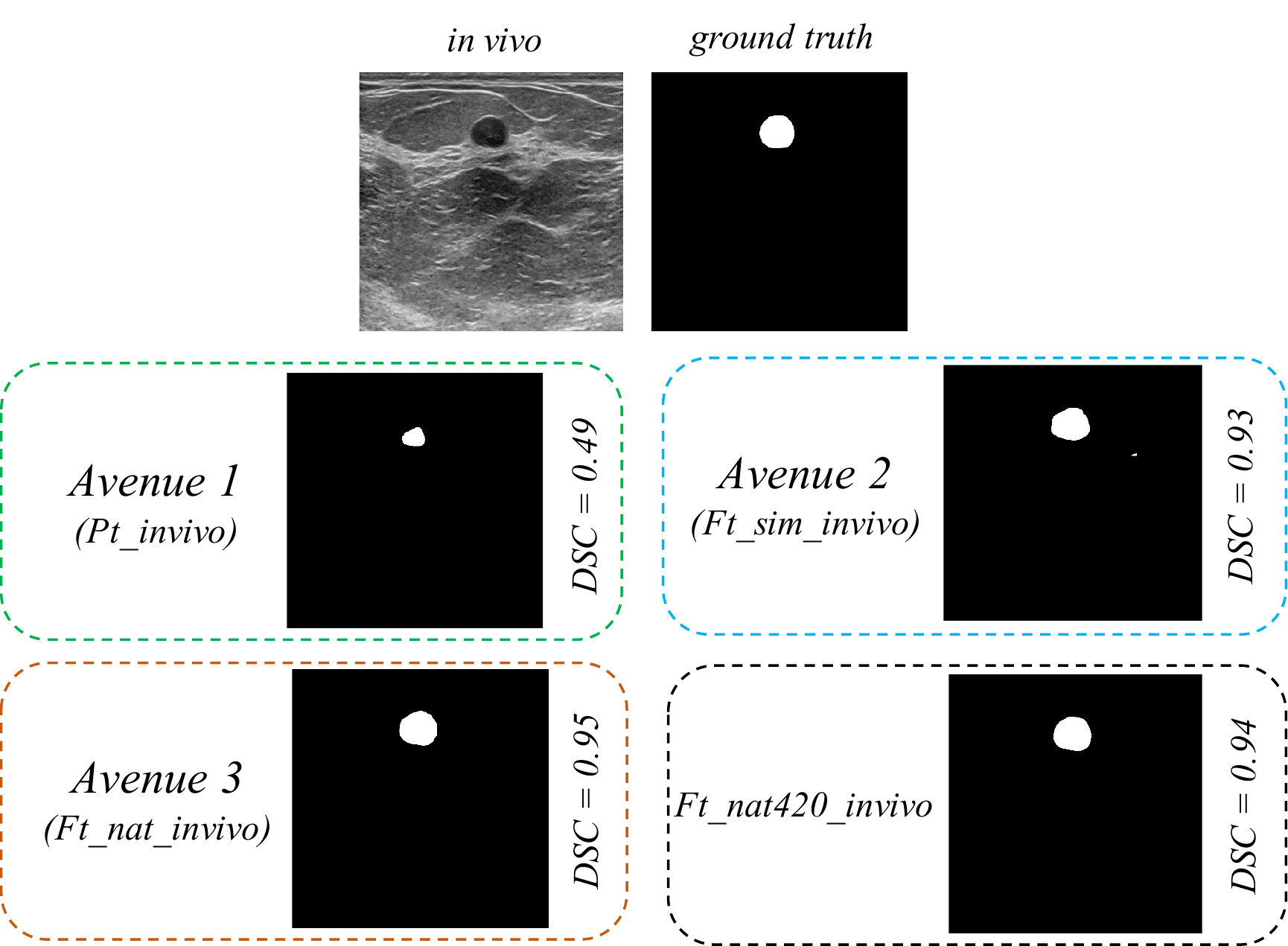}}
  \caption{Examples of segmentation results and their $DSC$ scores derived from $Avenue$ $1$, $Avenue$ $2$, $Avenue$ $3$, and $Ft\_nat420\_invivo$.}
  \label{fig:res}
\end{figure}

Table \ref{tab:dsc} presents the $DSC$ scores of the predicted masks derived from {\it Pt\_invivo}, \textit{Pt\_sim}, {\it Ft\_sim\_invivo}, \textit{Pt\_nat}, and {\it Ft\_nat\_invivo} networks for both training and testing \textit{in vivo} sets. The $DSC$ score for test set increases when we fine-tune the pre-trained network no matter what type of images were used during pre-training. Therefore, pre-training the network performs better than training from scratch with limited training data. It is worth mentioning that we used $6000$ number of natural images and $420$ number of simulated images during pre-training. However, when we decreased the number of natural images in $Avenue$ $3$ from $6000$ to $420$ in order to be equal to the number of simulated images, the $DSC$ score was reduced from $0.56$ to $0.38$ as shown in Table \ref{tab:dsc} (\textit{Pt\_nat420}, and {\it Ft\_nat420\_invivo} are referred as repetition of $Avenue$ $3$ using $420$ natural images). As a result, pre-training the network using simulated data is preferable as the auxiliary dataset than using natural images when same number of images from both datasets is available. Figure \ref{fig:res} demonstrates examples of the predicted masks with their $DSC$ scores.

We had $6000$ natural images in which $29$ hours was needed to train the {\it Ft\_nat\_invivo} network. For training on simulation, $2$ hours, and for training/fine-tuning on \textit{in vivo}, $5$ minutes was needed. As more annotations become available, although the U-Net is better trained, more time is needed for the pre-training step.

\begin{table}[h!]
 	\centering
	\caption{Mean and standard deviation of $DSC$ scores for predicted masks of \textit{in vivo} train and test sets over $5$-fold cross-validation}
	\label{tab:dsc}
	\small
 	\begin{tabular}{c l | r r}
         	\toprule[0.5mm]
             	\multicolumn{2}{c|}{\bf Network Name} & \multicolumn{1}{c}{ \textbf{Train \textit{in vivo}}} & \multicolumn{1}{c}{ \textbf{Test \textit{in vivo}}}\\
            	\toprule
         	\multirow{1}{*}{$Avenue$ $1$} & {\it Pt\_invivo} & $0.73 \pm 0.03$ & $0.37 \pm 0.04$ \\
            	\midrule
         	\multirow{2}{*}{$Avenue$ $2$} & {\it Pt\_sim} & $0.29 \pm 0.22$ & $0.27 \pm 0.19$\\
                 & {\it Ft\_sim\_invivo} & $0.79 \pm 0.05$ & $\bf 0.45 \pm 0.03$\\
         	\midrule
            	\multirow{2}{*}{$Avenue$ $3$} & {\it Pt\_nat} & $0.13 \pm 0.27$ & $0.14 \pm 0.26$\\
                 & {\it Ft\_nat\_invivo} & $0.85 \pm 0.04$ & $\bf 0.57 \pm 0.02$\\
         	\toprule
                 \multicolumn{2}{l|}{{\it Ft\_nat420\_invivo}} & $0.78 \pm 0.15$ & $0.40 \pm 0.03$\\
         	\toprule[0.5mm]
 	\end{tabular}
\end{table}

\section{Conclusion and future work}
\label{sec:con}
In this work, we showed that pre-training the network performs better than training the network from scratch especially when the number of annotations is limited. We proposed the use of simulated US images as the auxiliary dataset for pre-training the network. In addition, we confirmed that natural images can be also considered as the auxiliary dataset, however, thousands of them are required for optimum results which led to hours of pre-training. Therefore, we conclude that US simulation images are the preferred auxiliary dataset for pre-training the network. As our future work, we will validate our strategy for different type of {\it in vivo} datasets and segmentation applications such as prostate cancer and muscle segmentation.
\section{Acknowledgement}
This research was funded by Richard and Edith Strauss Foundation and by NSERC Discovery Grant RGPIN 04136. The authors would like to thank NVIDIA for donating the GPU.
\bibliographystyle{IEEEbib}
\bibliography{strings,refs}

\begin{thebibliography}{10}

\bibitem{ref2}
MH~Yap, G~Pons, J~Mart{\'\i}, S~Ganau, M~Sent{\'\i}s, R~Zwiggelaar, A~K
  Davison, and R~Mart{\'\i},
\newblock ``Automated breast ultrasound lesions detection using convolutional
  neural networks,''
\newblock {\em IEEE Journal of Biomedical and Health Informatics}, vol. 22, no.
  4, pp. 1218--1226, 2017.

\bibitem{ref10}
S~Liu, Y~Wang, X~Yang, B~Lei, L~Liu, SX~Li, D~Ni, and T~Wang,
\newblock ``Deep learning in medical ultrasound analysis: A review,''
\newblock {\em Engineering}, 2019.

\bibitem{ref1}
O~Ronneberger, P~Fischer, and T~Brox,
\newblock ``U-net: Convolutional networks for biomedical image segmentation,''
\newblock in {\em International Conference on Medical Image Computing and
  Computer-assisted Intervention}. Springer, 2015, pp. 234--241.

\bibitem{ref11}
S~Pereira, A~Pinto, V~Alves, and CA~Silva,
\newblock ``Brain tumor segmentation using convolutional neural networks in mri
  images,''
\newblock {\em IEEE Transactions on Medical Imaging}, vol. 35, no. 5, pp.
  1240--1251, 2016.

\bibitem{ref12}
HC~Shin, NA~Tenenholtz, JK~Rogers, CG~Schwarz, ML~Senjem, JL~Gunter,
  KP~Andriole, and M~Michalski,
\newblock ``Medical image synthesis for data augmentation and anonymization
  using generative adversarial networks,''
\newblock in {\em International Workshop on Simulation and Synthesis in Medical
  Imaging}. Springer, 2018, pp. 1--11.

\bibitem{ref3}
JA~Jensen,
\newblock ``Field: A program for simulating ultrasound systems,''
\newblock in {\em 10th Nordicbaltic Conference on Biomedical Imaging, Vol. 4,
  Supplement 1, Part 1: 351--353}. Citeseer, 1996.

\bibitem{ref8}
C~Xia, J~Li, X~Chen, A~Zheng, and Y~Zhang,
\newblock ``What is and what is not a salient object? learning salient object
  detector by ensembling linear exemplar regressors,''
\newblock in {\em Proceedings of the IEEE Conference on Computer Vision and
  Pattern Recognition}, 2017, pp. 4142--4150.

\bibitem{ref5}
V~Nair and GE~Hinton,
\newblock ``Rectified linear units improve restricted boltzmann machines,''
\newblock in {\em Proceedings of the 27th International Conference on Machine
  Learning (ICML-10)}, 2010, pp. 807--814.

\bibitem{ref6}
DP~Kingma and J~Ba,
\newblock ``Adam: A method for stochastic optimization,''
\newblock {\em arXiv preprint arXiv:1412.6980}, 2014.

\bibitem{ref7}
K~He, X~Zhang, S~Ren, and J~Sun,
\newblock ``Delving deep into rectifiers: Surpassing human-level performance on
  imagenet classification,''
\newblock in {\em Proceedings of the IEEE International Conference on Computer
  Vision}, 2015, pp. 1026--1034.

\end{thebibliography}

\end{document}